# Influence of surface plasmon polaritons on laser energy absorption and structuring of surfaces

P. N. Terekhin (1 and 2)[*], O. Benhayoun (3), S. T. Weber (1), D. S. Ivanov (1, 3 and 4), M. E. Garcia (3) and B. Rethfeld (1)

((1) Technische Universität Kaiserslautern, Kaiserslautern, Germany, (2) National Research Centre "Kurchatov Institute", Moscow, Russia, (3) University of Kassel, Kassel, Germany, (4) Lebedev Physical Institute, Moscow, Russia)

The accurate calculation of laser energy absorption during femto- or picosecond laser pulse experiments is very important for the description of the formation of periodic surface structures. On a rough material surface, a crack or a step edge, ultrashort laser pulses can excite surface plasmon polaritons (SPP), i.e. surface plasmons coupled to a laser-electromagnetic wave. The interference of such plasmon wave and the incoming pulse leads to a periodic modulation of the deposited laser energy on the surface of the sample. In the present work, within the frames of a Two Temperature Model we propose the analytical form of the source term, which takes into account SPP excited at a step edge of a dielectric-metal interface upon irradiation of an ultrashort laser pulse at normal incidence. The influence of the laser pulse parameters on energy absorption is quantified for the example of gold. This result can be used for nanophotonic applications and for the theoretical investigation of the evolution of electronic and lattice temperatures and, therefore, of the formation of surfaces with predestined properties under controlled conditions.

*E-mail address: terekhin@physik.uni-kl.de (P. N. Terekhin)

## 1. Introduction

The mechanisms of laser-matter interaction in condensed matter are of great importance for many applications [1-5] including development of promising ultrafast nanophotonic devices [6-11]. Laser-induced surface phenomena could be produced by pre-structured beams [12] or excitation of surface plasmon polariton (SPP) waves [13,14], i.e. surface plasmons coupled to an incident laser pulse. To excite SPP waves, special conditions are needed, for example, prism or grating coupling [13], surface roughness [14] or a step edge of a sample [15,16]. The interference of the SPP and laser electromagnetic fields results in a periodic profile of a deposited laser energy along a surface of a solid. One of the possible scenarios for the description of laser-induced periodic surface structures (LIPSS) is based on the SPP excitation and their interference with the incident beam [17-19]. LIPSS patterns could appear on the surface after either single [20 - 22] or multiple [19,23] laser shots. It was shown in a number of papers, that these structures can be produced on different types of materials, metals, semiconductors, dielectrics [19 - 25], and for variety of incident angles [26].

In Ref. [22] for gold and in Ref. [27] for titanium and silicon the authors investigated the evolution of electron and lattice temperatures using the two-temperature model (TTM) [28], where the laser energy absorption was implemented by an arbitrary periodic function, which was assumed to arise from the interference of the SPP and incident light. The laser energy source for Au and $SiO_2$ was described in the Ref. [29] taking into account the changing of the optical properties of the irradiated sample during the pulse duration, but the periodicity along the direction parallel to the surface was also introduced as an arbitrary cosine-type perturbation.

In this paper, we investigate the energy absorption upon irradiation of solid targets by ultrashort laser pulses. An analytical function for the source term in the frames of TTM approach is derived by explicit calculation of the interference of the SPP fields and the laser fields. This source term could further be used for the investigation of laser matter interaction and the formation of surface structures [12,30]. One can also apply our source function for studying the properties of the excited SPP in the framework of scanning near-field optical microscopy (SNOM) or of two-photon photoemission electron microscopy (2P-PEEM).

**2. Theory**

It is well known that SPP excitation is not possible on a smooth surface, since SPP and light dispersion relations do not intersect [13,14]. However, SPP can be excited, for example, in the presence of roughness, grating structures or at a step edge [13,14] even with a single pulse [20 - 22]. We present results for the excitation of the SPP at a step edge of a metal sample and a dielectric media. It has been shown in Ref. [15] that the SPP are excited at the position of the step edge even if we shine to the entire surface. For our calculations the laser pulse is chosen to be normal to the surface. The considered irradiation geometry is presented in Fig. 1. The origin of the coordinate system is at the step edge.

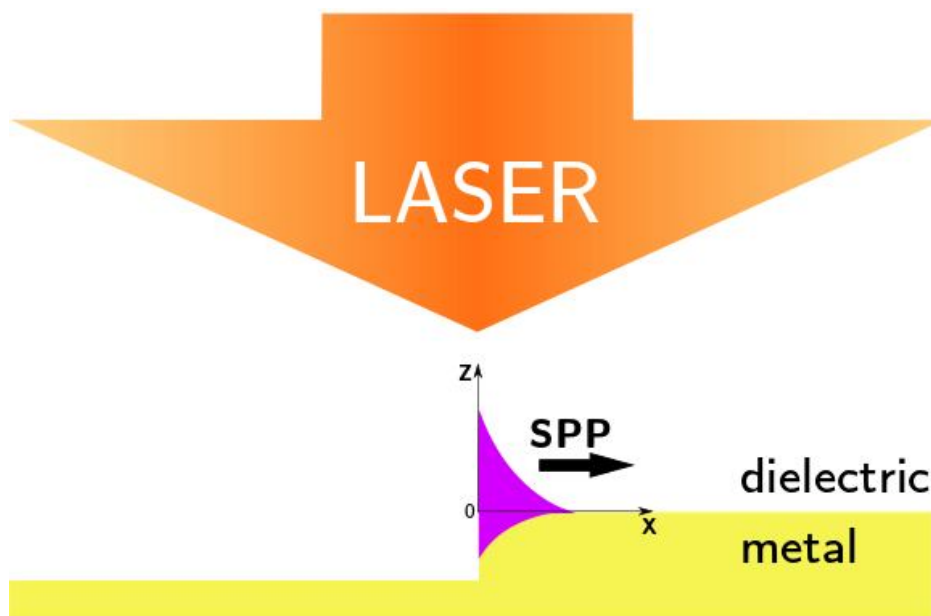

**Figure 1.** The scheme of the irradiation geometry with the SPP excitation.

We suppose the relative permeability of the nonmagnetic sample $\mu = 1$ and consider incident light with electric $\mathbf{E}_{\text{las}}^{\text{inc}} = \left(E_{\text{las,x}}^{\text{inc}}; 0; 0\right)$ and magnetic $\mathbf{H}_{\text{las}}^{\text{inc}} = \left(0; H_{\text{las,y}}^{\text{inc}}; 0\right)$ fields described by a Gaussian pulse shape of the form

$$E_{\text{las,x}}^{\text{inc}} = E_0 e^{-ik_0 z} e^{-i\omega t} \xi_{\text{las}}(x, y, t), \tag{1}$$

$$H_{\text{las,y}}^{\text{inc}} = H_0 e^{-ik_0 z} e^{-i\omega t} \xi_{\text{las}}(x, y, t), \tag{2}$$

where $E_0$ is the amplitude of the incident electric field, $H_0 = -c\varepsilon_0 E_0$ is the amplitude of the incident magnetic field, $k_0 = \omega/c$ is the wave vector of light, $\omega$ is the laser angular frequency, $c$ is the speed of light, $\varepsilon_0$ is the permittivity in vacuum and

$$\xi_{\text{las}}(x, y, t) = \frac{1}{d_{\text{las}}}\sqrt{\frac{\sigma}{\pi}} e^{-\sigma\frac{x^2+y^2}{2d_{\text{las}}^2}} \sqrt{\frac{1}{\tau}\sqrt{\frac{\sigma}{\pi}}} e^{-\sigma\frac{(t-t_0)^2}{2\tau^2}}, \tag{3}$$

with $\sigma = 4\ln 2$, $d_{\text{las}}$ is the focus diameter at FWHM (full width at half maximum), $\tau$ is a pulse duration at FWHM, $t_0 = 3\tau$ is chosen as a location of the pulse maximum.

The reflected fields $\mathbf{E}_{\text{las}}^{\text{r}} = \left(E_{\text{las,x}}^{\text{r}}; 0; 0\right)$ and $\mathbf{H}_{\text{las}}^{\text{r}} = \left(0; H_{\text{las,y}}^{\text{r}}; 0\right)$ are given by

$$E_{\text{las,x}}^{\text{r}} = E_0^{\text{r}} e^{ik_0 z} e^{-i\omega t} \xi_{\text{las}}(x, y, t), \tag{4}$$

$$H_{\text{las,y}}^{\text{r}} = H_0^{\text{r}} e^{ik_0 z} e^{-i\omega t} \xi_{\text{las}}(x, y, t), \tag{5}$$

where $E_0^{\text{r}}$ is the amplitude of the reflected electric field, $H_0^{\text{r}} = c\varepsilon_0 E_0^{\text{r}}$ is the amplitude of the reflected magnetic field. Finally, the absorbed fields $\mathbf{E}_{\text{las}}^{\text{a}} = \left(E_{\text{las,x}}^{\text{a}}; 0; 0\right)$ and $\mathbf{H}_{\text{las}}^{\text{a}} = \left(0; H_{\text{las,y}}^{\text{a}}; 0\right)$ are given by

$$E_{\text{las,x}}^{\text{a}} = E_0^{\text{a}} e^{-ik_0 \tilde{n}_m z} e^{-i\omega t} \xi_{\text{las}}(x, y, t), \tag{6}$$

$$H_{\text{las,y}}^{\text{a}} = H_0^{\text{a}} e^{-ik_0 \tilde{n}_m z} e^{-i\omega t} \xi_{\text{las}}(x, y, t), \tag{7}$$

where $E_0^{\text{a}}$ is the amplitude of the absorbed electric field, $H_0^{\text{a}} = -c\varepsilon_0 E_0^{\text{a}}$ is the amplitude of the absorbed magnetic field and $\tilde{n}_m$ is the complex refractive index of a metal defined as

$$\tilde{n}_m(\omega) = n_m(\omega) + ik_m(\omega). \tag{8}$$

In order to determine electromagnetic field produced by the SPP, we are using the following expressions $\mathbf{E}_{SPP}^{r} = \left(E_{SPP,x}^{r}; 0; E_{SPP,z}^{r}\right)$ and $\mathbf{H}_{SPP}^{r} = \left(0; H_{SPP,y}^{r}; 0\right)$, which are based on the solution of Maxwell's equations [13], where we assume a Gaussian pulse profile in the following form for z > 0 (dielectric)

$$H_{SPP,y}^{r} = A_1 e^{ik_x x} e^{-k_{z,d} z} e^{-i\omega t} \xi_{SPP}(x,y,t), \tag{9}$$

$$E_{SPP,x}^{r} = iA_1 \frac{k_{z,d}}{\omega \varepsilon_0 \varepsilon_d} e^{ik_x x} e^{-k_{z,d} z} e^{-i\omega t} \xi_{SPP}(x,y,t), \tag{10}$$

$$E_{SPP,z}^{r} = -A_1 \frac{k_x}{\omega \varepsilon_0 \varepsilon_d} e^{ik_x x} e^{-k_{z,d} z} e^{-i\omega t} \xi_{SPP}(x,y,t), \tag{11}$$

where $A_1$ is the amplitude of the magnetic field of the SPP in the dielectric media, $k_x = k_x' + ik_x''$ is the wave vector of the SPP in the direction of propagation; $k_{z,d}$ is the wave vector of the SPP in dielectric in the direction perpendicular to the interface between two media, $\varepsilon_d$ is the dielectric function of the dielectric. The function $\xi_{SPP}(x,y,t)$ can be written as [6,15,16]

$$\xi_{SPP}(x,y,t) = \frac{1}{d_{las}} \sqrt{\frac{\sigma}{\pi}} e^{-\sigma \frac{y^2}{2d_{las}^2}} \sqrt{\frac{1}{\tau} \sqrt{\frac{\sigma}{\pi}}} e^{-\sigma \frac{\left(x - v_{g,SPP}(t-t_0)\right)^2}{2v_{g,SPP}^2 \tau^2}}, \tag{12}$$

where $v_{g,SPP} = d\omega / dk_x'$ is the group velocity of the SPP.

The expressions $\mathbf{E}_{SPP}^{a} = \left(E_{SPP,x}^{a}; 0; E_{SPP,z}^{a}\right)$ and $\mathbf{H}_{SPP}^{a} = \left(0; H_{SPP,y}^{a}; 0\right)$ of the SPP fields in the half-space z < 0 (metal) are

$$H_{SPP,y}^{a} = A_2 e^{ik_x x} e^{k_{z,m} z} e^{-i\omega t} \xi_{SPP}(x,y,t), \tag{13}$$

$$E_{SPP,x}^{a} = -iA_2 \frac{k_{z,m}}{\omega \varepsilon_0 \varepsilon_m} e^{ik_x x} e^{k_{z,m} z} e^{-i\omega t} \xi_{SPP}(x,y,t), \tag{14}$$

$$E_{SPP,z}^{a} = -A_2 \frac{k_x}{\omega \varepsilon_0 \varepsilon_m} e^{ik_x x} e^{k_{z,m} z} e^{-i\omega t} \xi_{SPP}(x,y,t), \tag{15}$$

where $A_2$ is the amplitude of the magnetic field of the SPP in the metal, $k_{z,m}$ is the wave vector of the SPP in the metal in the direction perpendicular to the interface between two media, $\varepsilon_m = \varepsilon'_m + i\varepsilon''_m = \tilde{n}_m^2$ is the dielectric function of the metal. The SPP fields should fulfill Maxwell's equations, yielding [13]

$$k_{z,d}^2 = k_x^2 - k_0^2 \varepsilon_d, \qquad (16)$$

$$k_{z,m}^2 = k_x^2 - k_0^2 \varepsilon_m. \qquad (17)$$

Thus, we have in the dielectric media incident and reflected laser fields and SPP fields, whereas in the metal we have absorbed laser fields and corresponding SPP fields. In accordance with the boundary conditions of Maxwell's equations at the interface between two media, we find respective amplitudes and relation between $k_{z,d}$ and $k_{z,m}$:

$$E_0^r = \frac{1-\tilde{n}_m}{1+\tilde{n}_m} E_0, \qquad (18)$$

$$E_0^a = \frac{2}{1+\tilde{n}_m} E_0, \qquad (19)$$

$$A_1 = A_2, \qquad (20)$$

$$\frac{k_{z,d}}{\varepsilon_d} + \frac{k_{z,m}}{\varepsilon_m} = 0. \qquad (21)$$

We associate the amplitude of the SPP magnetic field $A_1$ with the amplitude of the incident magnetic field of a laser

$$A_1 = |H_0|\beta e^{i\delta}, \qquad (22)$$

where $\beta$ is the coupling efficiency and $\delta$ is the phase difference between the incident beam and the excited SPP, which we are leaving the only free parameters.

Combining Eq. (21) with Eqs. (16)-(17) we define a dispersion relation for the SPP as [13]

$$k_x = k_0 \sqrt{\frac{\varepsilon_m \varepsilon_d}{\varepsilon_m + \varepsilon_d}}, \qquad (22)$$

We assume that the dielectric functions do not depend on space and time. In this case, we can express the absorption of the laser energy as

$$Q_{\text{total}}(\mathbf{r},t) = -div(\mathbf{S}_{\text{total}}(\mathbf{r},t)), \qquad (23)$$

where $\mathbf{S}_{\text{total}}$ is the Poynting vector giving the laser depositing energy per unit time and per unit area. We will call $Q_{\text{total}}(\mathbf{r},t)$ the source term entering TTM from the perspective of the electronic subsystem of the metal. We use SPP fields (13)-(15) and laser fields (6)-(7) to calculate the Poynting vector

$$\mathbf{S}_{\text{total}}(\mathbf{r},t) = \mathbf{E}_{\text{total}} \times \mathbf{H}_{\text{total}}, \qquad (24)$$

where

$$\mathbf{E}_{\text{total}} = \mathbf{E}_{\text{las}}^{a} + \mathbf{E}_{\text{SPP}}^{a}, \qquad (25)$$

$$\mathbf{H}_{\text{total}} = \mathbf{H}_{\text{las}}^{a} + \mathbf{H}_{\text{SPP}}^{a}. \qquad (26)$$

Averaging the Poynting vector over one full period of the laser pulse $2\pi/\omega$, we obtain an expression for the source term

$$Q_{\text{total}}(\mathbf{r},t) = Q_{\text{las-las}}(\mathbf{r},t) + Q_{\text{las-SPP}}(\mathbf{r},t) + Q_{\text{SPP-SPP}}(\mathbf{r},t), \qquad (27)$$

where the first term

$$Q_{\text{las-las}}(\mathbf{r},t) = F_{\text{inc}} \frac{8 k_0 n_m k_m}{(n_m+1)^2 + k_m^2} \xi_{\text{las}}^2 \exp(2 k_0 k_m z) \qquad (28)$$

arise from the laser-laser interference, where $F_{\text{inc}}$ is the incident laser fluence. In case of $\beta = 0$ (no SPP) it gives automatically the common expression for the source term [5,12]

$$Q_{\text{total}}^{\beta=0}(\mathbf{r},t) = Q_{\text{las-las}}(\mathbf{r},t) = F_{\text{inc}} \alpha_{\text{abs}} (1-R) e^{\alpha_{\text{abs}} z} \frac{\sigma}{\pi d_{\text{las}}^2} e^{-\sigma \frac{x^2+y^2}{d_{\text{las}}^2}} \frac{1}{\tau}\sqrt{\frac{\sigma}{\pi}} e^{-\sigma \frac{(t-t_0)^2}{\tau^2}}, \qquad (29)$$

where the absorption coefficient $\alpha_{\text{abs}}$ was defined as

$$\alpha_{\text{abs}} = 2 k_0 k_m \qquad (30)$$

and reflectivity

$$R = \frac{(n_m-1)^2 + k_m^2}{(n_m+1)^2 + k_m^2}. \qquad (31)$$

The second term describes laser-SPP interference:

$$Q_{\text{las-SPP}}(\mathbf{r},t) = \beta F_{\text{inc}} \frac{2\xi_{\text{las}}\xi_{\text{SPP}}}{(n_m+1)^2 + k_m^2} \{f_1 \cos(f_3) + f_2 \sin(f_3)\} \exp\left((k_0 k_m + k'_{z,m})z - k''_x x\right), \quad (32)$$

where values $f_1 - f_7$ are given in the Appendix A.

The periodicity along the lateral distance $x$ in $Q_{\text{total}}$ comes from this term. The SPP wavelength is defined from Eqs. (32) as (see also the Eq. (A3) in the Appendix A)

$$\lambda_{\text{SPP}} = \frac{2\pi}{k'_x}. \quad (33)$$

The third term, which describes propagating SPP, is a result of the SPP-SPP interference:

$$Q_{\text{SPP-SPP}}(\mathbf{r},t) = \beta^2 F_{\text{inc}} \frac{2\left[\varepsilon'_m(k'_x k''_x - k'_{z,m} k''_{z,m}) + \varepsilon''_m(k'^2_x + k'^2_{z,m})\right]}{k_0 |\varepsilon_m|^2} \xi^2_{\text{SPP}} \exp\left(2(k'_{z,m}z - k''_x x)\right). \quad (34)$$

Finally, we should mention, that our analytical source term can be applied for different materials irradiated by ultrashort laser pulses with arbitrary beam parameters.

## 3. Results and discussion

To demonstrate the ability of our source term to describe real experimental situations, we have studied the irradiation of a gold sample in surrounding air. For the dielectric function of gold, the experimental data were taken from the Ref. [7] (therein the case of the evaporated gold sample) while for air we set $\varepsilon_d = 1$. We have used the laser parameters of $F_{\text{inc}} = 0.1 \text{ J/cm}^2$, $d_{\text{las}} = 70\,\mu\text{m}$, $\tau = 40\,\text{fs}$, $t_0 = 3\tau$. For the given laser wavelength the characteristics of the source term, such as the SPP wavelength $\lambda_{\text{SPP}}$, the propagation length $L^x_{\text{SPP}} = 1/(2k''_x)$, the SPP penetration depth $L^z_{\text{SPP},m} = 1/(2k'_{z,m})$ and the laser penetration depth $L^z_{\text{laser}} = 1/\alpha_{\text{abs}}$, are calculated. In order to investigate the relation between the SPP wavelength, see Eq. (33), and the laser wavelength, we link them through the definition of the phase velocity of the SPP

$$v_{\text{ph, SPP}} = \omega / k'_x \quad (35)$$

in the form

$$\lambda_{\text{SPP}} = \frac{v_{\text{ph, SPP}}(\lambda)}{c}\lambda = \frac{\omega}{ck'_x(\lambda)}\lambda. \tag{36}$$

Fig. 2 shows the SPP wavelength $\lambda_{\text{SPP}}$ according to the Eq. (36) divided by the laser wavelength $\lambda$, i.e. $\lambda_{\text{SPP}}/\lambda$ in dependence on the laser wavelength (left axis). Additionally, the absorption coefficient $\alpha_{\text{abs}}$ of gold, see Eq. (30), is shown (right axis). The experimental data from Ref. [7] were used.

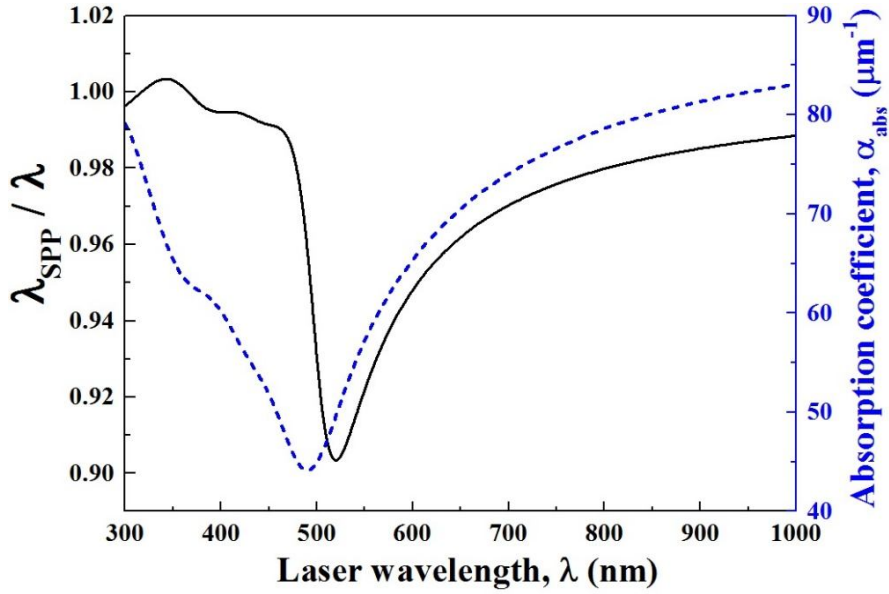

**Figure 2.** Dependence of the wavelength ratio $\lambda_{\text{SPP}}/\lambda$ and the absorption coefficient $\alpha_{\text{abs}}$ on the the laser wavelength $\lambda$.

It is seen from the Fig. 2 that generally the SPP wavelength $\lambda_{\text{SPP}}$ is lower than the laser wavelength $\lambda$ for the given range. The dip in the ratio of the wavelength reflects the processes of photo-excitation of d electrons of gold [31,32]. The dip in the absorption coefficient is also related to the d electrons, but the position of the peak is slightly shifted, because the SPP wavelength $\lambda_{\text{SPP}}$ and the absorption coefficient have different dependencies on the dielectric function (compare Eq. (30) and Eq. (33)).

The dependence of the propagation length $L^x_{\text{SPP}}$ on the laser wavelength $\lambda$ is presented in Fig. 3.

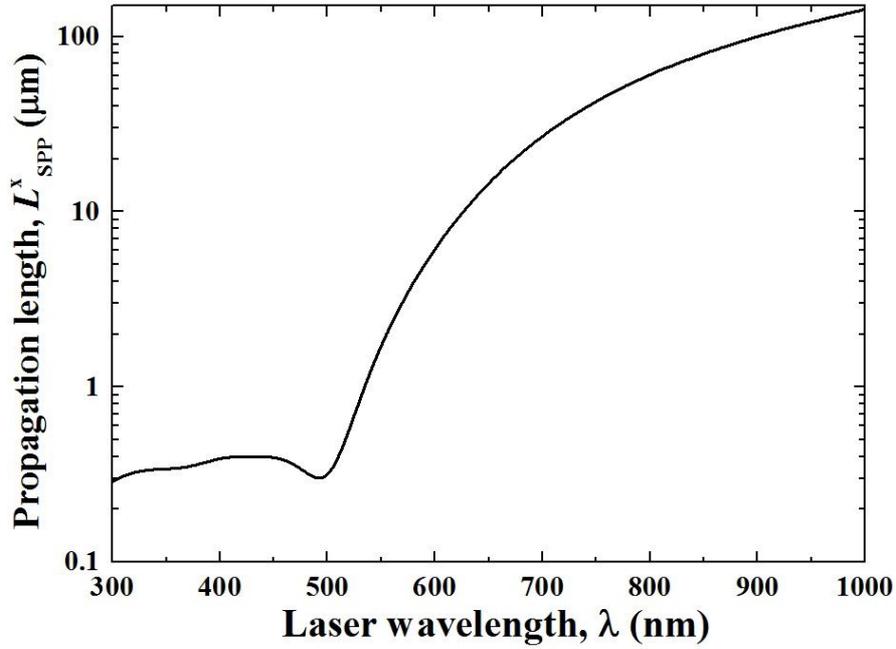

**Figure 3.** Dependence of the propagation length $L_{SPP}^x$ on the the laser wavelength $\lambda$.

Fig. 3 shows that the SPP decay quickly for wavelengths shorter than 500 nm, but propagate far from the position of excitation in case of increasing laser wavelength. For the following demonstration of the application of our source term, we have chosen wavelengths from two different regions, namely 800 nm and 400 nm. The characteristic lengths for these wavelengths are presented in Table 1.

| Laser wavelength (nm) | $\lambda_{SPP}$ (nm) | $L_{SPP}^x$ (μm) | $L_{SPP,m}^z$ (nm) | $L_{laser}^z$ (nm) |
|---|---|---|---|---|
| 800 | 784 | 60 | 12.46 | 12.72 |
| 400 | 398 | 0.39 | 15.49 | 16.62 |

**Table 1.** Calculated optical parameters of gold.

It is clearly seen from Table 1, that the propagation length $L_{SPP}^x$ at the wavelength of $\lambda = 400\,\text{nm}$ is much smaller than the one of 800 nm, but the penetration depth of the SPP $L_{SPP,m}^z$ and the laser penetration depths $L_{laser}^z$ are a little bit larger for the lower wavelength.

Fig. 4 shows the source term $Q_{total}$ at the surface and at the maximum intensity of the laser pulse according to the Eq. (27) including the influence of the SPP normalized to the source term of the laser only, i.e. $Q_{total}/Q_{las-las}$, in dependence on the lateral distance from the step edge of x=0. We set the coupling efficiency to $\beta = 0.8$ and the phase shift to $\delta = 0$ for the best representation of the typical observations. The dependence of the source term ratio on the lateral distance from the step edge at different coupling efficiencies $\beta$ and the phases $\delta$ is given in the Appendix B.

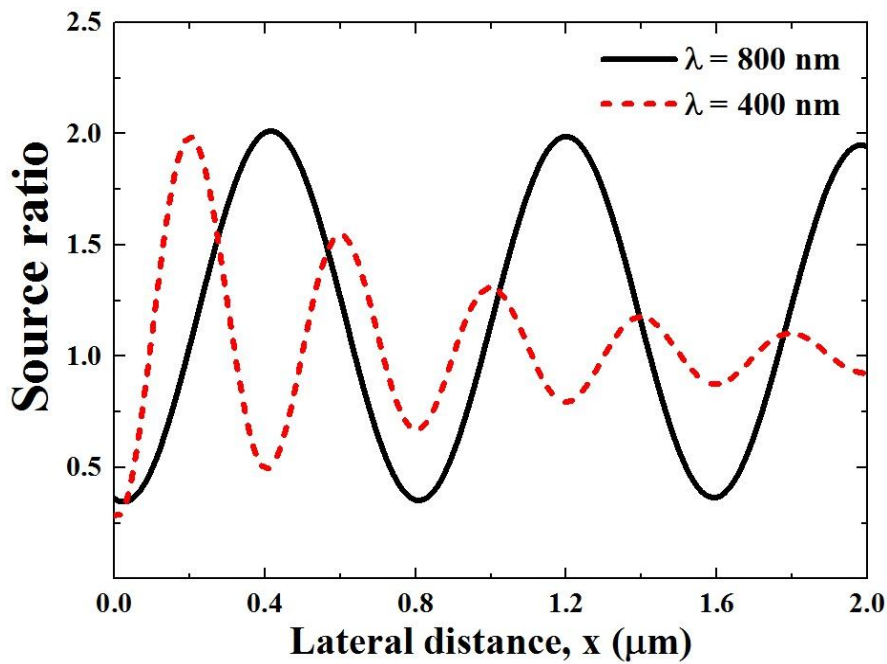

**Figure 4.** Dependence of the source term ratio on the lateral distance from the step edge at different laser wavelengths of $\lambda = 800\,\text{nm}$ and $\lambda = 400\,\text{nm}$ at the surface $(z=0, y=0)$ and the moment of the pulse maximum $t = t_0$.

We can see from the Fig. 4 the modulation of the absorbed energy with the periods $\lambda_{SPP}$ given in Table 1. We can also observe the enhancement of the source term, because it takes into account excited SPP and as a result we have two more terms in the Eq. (27). In contrast to the SPP excitation with the 800 nm laser wavelength, the SPP excited at 400 nm decay quickly and do not propagate along the sample surface. It means that the formation of periodic surface structures could be possible with smaller periodicity if we decrease the laser wavelength, but the area of the sample, where these structures could be created, is limited by the decay length of the SPP.

In order to reveal further features of the absorbed energy distribution, Fig. 5 presents the dependence of $Q_{total}(\mathbf{r},t)$ on time and on lateral distance at the surface of gold for the wavelength of 800 nm.

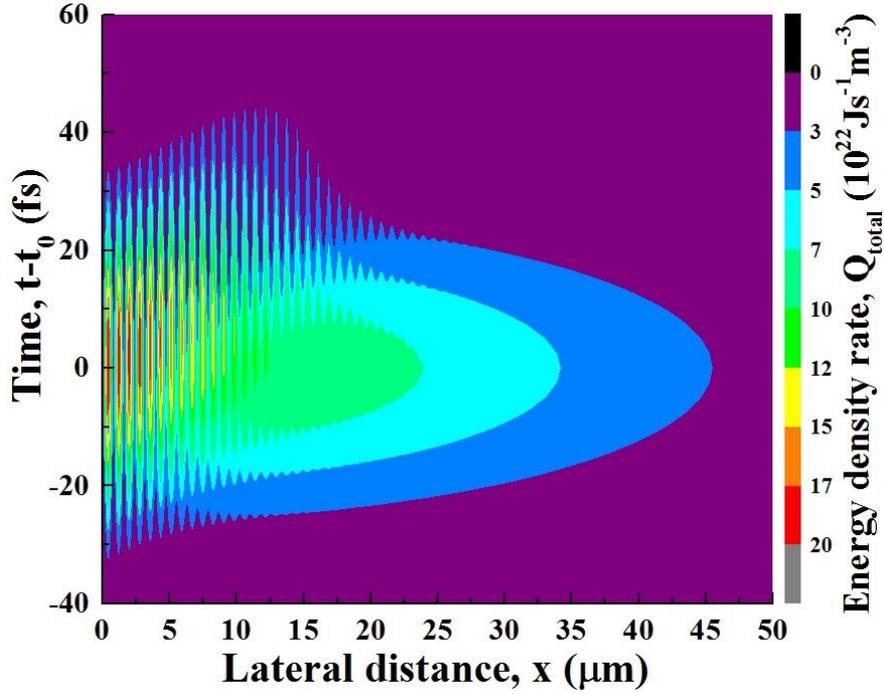

**Figure 5.** The energy absorbed by the electrons of gold $Q_{total}(x,t)$ at the surface for the laser wavelength of $\lambda = 800\,\text{nm}$ and the pulse duration of $\tau = 40\,\text{fs}$.

We can observe from the Fig. 5 that the laser pulse excites the SPP, interferes with the SPP creating the periodic pattern shown in Fig. 4. The highest enhancement of the absorbed energy is near to the step edge at which the SPP are excited. After excitation the SPP propagate and decay along the sample and interfere with the laser until the laser pulse is over. The third term in Eq. (27) describes the SPP propagation after the pulse. We can see the signature of this propagating SPP as the hump at the top part on the Fig. 5. The shape of this hump will depend on the coupling efficiency $\beta$, that is why this value could be estimated from experiments.

The dependence of the source term on time and lateral distance at the surface for wavelength of $\lambda = 400\,\text{nm}$ is displayed in Fig. 6. Note the different scaling of the lateral direction as compared to Fig. 5, due to the smaller wavelength of the SPP $\lambda_{SPP}$. Apart from the periodic absorption of the deposited energy, it is seen from the Fig. 6 that the SPP decay quickly without formation of the hump like on the Fig. 5.

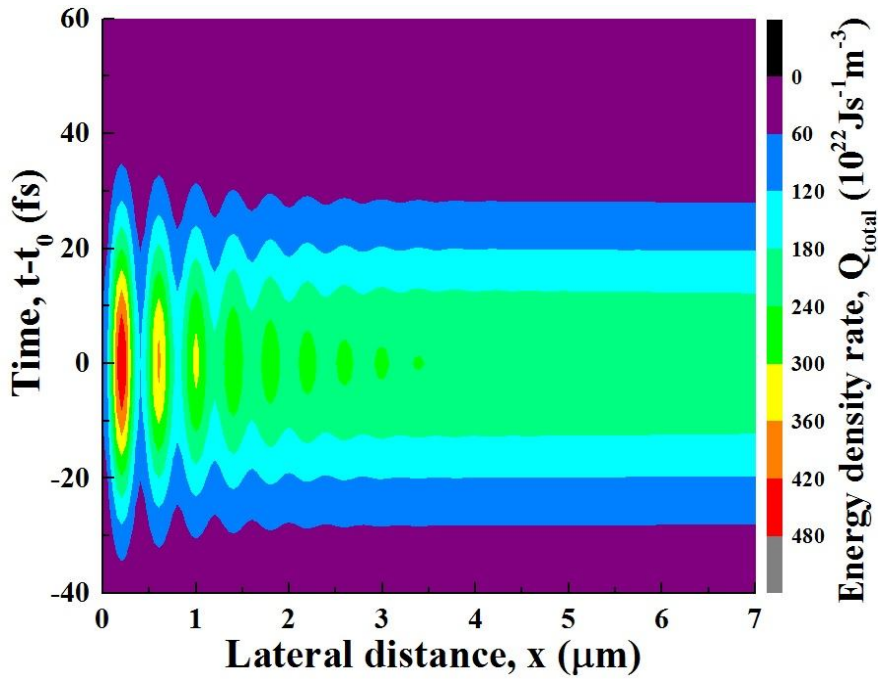

**Figure 6.** The energy absorbed by the electrons of gold $Q_{total}(x,t)$ at the surface for the laser wavelength of $\lambda = 400\,\text{nm}$ and the pulse duration of $\tau = 40\,\text{fs}$ reveals shorter periodicity and quicker decay in comparison with the result presented in the Fig. 4 with the laser wavelength of $\lambda = 800\,\text{nm}$.

The dependence of the source term on $\tau$ was investigated since it can be used for arbitrary laser pulse durations. Fig. 7 shows the time and space dependence of the absorbed energy at the surface in case of longer pulse duration.

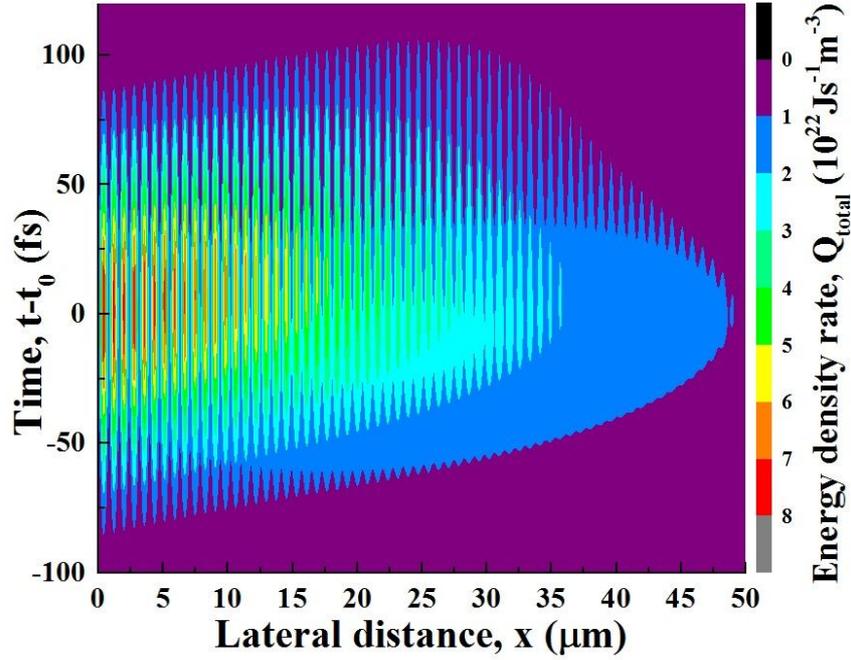

**Figure 7.** The energy absorbed by the electrons of gold $Q_{total}(x,t)$ at the surface for the same than in Fig. 5 laser wavelength of $\lambda = 800\,\text{nm}$ and the longer pulse duration of $\tau = 100\,\text{fs}$ displays the interference pattern along the longer lateral distance of the irradiated sample.

Fig. 7 depicts the periodic structure with the period mentioned in Table 1 for this wavelength. In contrast to the short pulse duration of Fig. 5, we can observe on the Fig. 7, that laser and SPP fields coexist and interfere longer time and along longer lateral distance. It results in the different hump shape formation in the top part on the Fig. 7 in comparison with the Fig. 5. This signature of the propagating SPP will also depend on the coupling efficiency $\beta$, which could be estimated from the experiments for different pulse durations.

## 4. Conclusions

The process of laser energy absorption has been investigated upon the excitation of surface plasmon polaritons under normal incidence irradiation of a metal sample with a step edge geometry. The interference of the SPP with the laser fields results in a periodic profile of the deposited energy along the surface. An analytical source term in the frames of a TTM approach has been derived, which consists of three parts arises from the laser- laser, laser-SPP and SPP-SPP interferences.

For the case of a gold sample in air, we have studied the influence of laser pulse parameters on the deposited energy distribution. The simulation results demonstrate a periodic modification and an enhancement of the absorbed energy for the two example laser wavelengths of 800 nm and 400 nm. In the latter case the SPP decay very fast and do not survive after the laser pulse. If we increase the duration of the pulse, the SPP and laser fields coexist on a longer time scale. Our result can be applied for describing different materials and arbitrary parameters of the laser pulses. Note that the present source term can be used for studying features of the excited SPP, for example, in the 2P-PEEM and SNOM experiments. It could help to shed light on the properties of plasmon-induced hot carriers for future nano-photonic applications.

**Acknowledgements**

We acknowledge the financial support of the Deutsche Forschungsgemeinschaft projects RE 1141/14-2, DFG IV 122/4-1, GA 465/15-2 and GA 465/18-1.

We are grateful to M. Hartelt for fruitful discussions. Additionally, the authors appreciate the Allianz für Hochleistungsrechnen Rheinland-Pfalz for providing computing resources through project MULAN on the Elwetritsch high performance computing cluster and the federal collective usage center, the Complex for Simulation and Data Processing for Mega-science Facilities at NRC 'Kurchatov Institute' (ministry subvention under agreement RFMEFI62117X0016), http://ckp.nrcki.ru/. Some calculations were performed at Lichtenberg Super Computer Facility TU-Darmstadt (Germany).

**Appendix A. Laser-SPP interference**

The values in the second term of the energy absorption rate Eq. (32) are the following:

$$f_1 = \frac{-k_x'' f_4 + k_x' f_5}{k_0 |\varepsilon_m|^2} - \left(k_{z,m}' + k_0 k_m\right) f_6 - \left(k_{z,m}'' + k_0 n_m\right) f_7, \tag{A1}$$

$$f_2 = -\frac{k_x'' f_5 + k_x' f_4}{k_0 |\varepsilon_m|^2} - \left(k_{z,m}' + k_0 k_m\right) f_7 + \left(k_{z,m}'' + k_0 n_m\right) f_6, \tag{A2}$$

$$f_3 = k_x' x + \left(k_{z,m}'' + k_0 n_m\right) z + \delta, \tag{A3}$$

$$f_4 = \varepsilon_m' \left[ k_x' \left(n_m + n_m^2 + k_m^2\right) + k_x'' k_m \right] + \varepsilon_m'' \left[ -k_x' k_m + k_x'' \left(n_m + n_m^2 + k_m^2\right) \right], \tag{A4}$$

$$f_5 = \varepsilon_m' \left[ k_x' k_m - k_x'' \left(n_m + n_m^2 + k_m^2\right) \right] + \varepsilon_m'' \left[ k_x' \left(n_m + n_m^2 + k_m^2\right) + k_x'' k_m \right], \tag{A5}$$

$$f_6 = 1 + n_m + \frac{\varepsilon_m' \left[ k_{z,m}' k_m - k_{z,m}'' \left(n_m + n_m^2 + k_m^2\right) \right]}{k_0 |\varepsilon_m|^2} + \\ + \frac{\varepsilon_m'' \left[ k_{z,m}' \left(n_m + n_m^2 + k_m^2\right) + k_{z,m}'' k_m \right]}{k_0 |\varepsilon_m|^2}, \tag{A6}$$

$$f_7 = -k_m - \frac{\varepsilon_m' \left[ k_{z,m}' \left(n_m + n_m^2 + k_m^2\right) + k_{z,m}'' k_m \right]}{k_0 |\varepsilon_m|^2} + \\ + \frac{\varepsilon_m'' \left[ k_{z,m}' k_m - k_{z,m}'' \left(n_m + n_m^2 + k_m^2\right) \right]}{k_0 |\varepsilon_m|^2}. \tag{A7}$$

**Appendix B. Dependence of the source term on the coupling efficiency $\beta$ and the phase $\delta$.**

Fig. B1 shows the source term $Q_{\text{total}}$ at the surface and at the maximum of the laser pulse according to the Eq. (27) including the influence of the SPP normalized to the source term of the laser only, i.e. $Q_{\text{total}} / Q_{\text{las-las}}$, in dependence on the lateral distance from the step edge of x=0 for the laser wavelength of 800 nm.

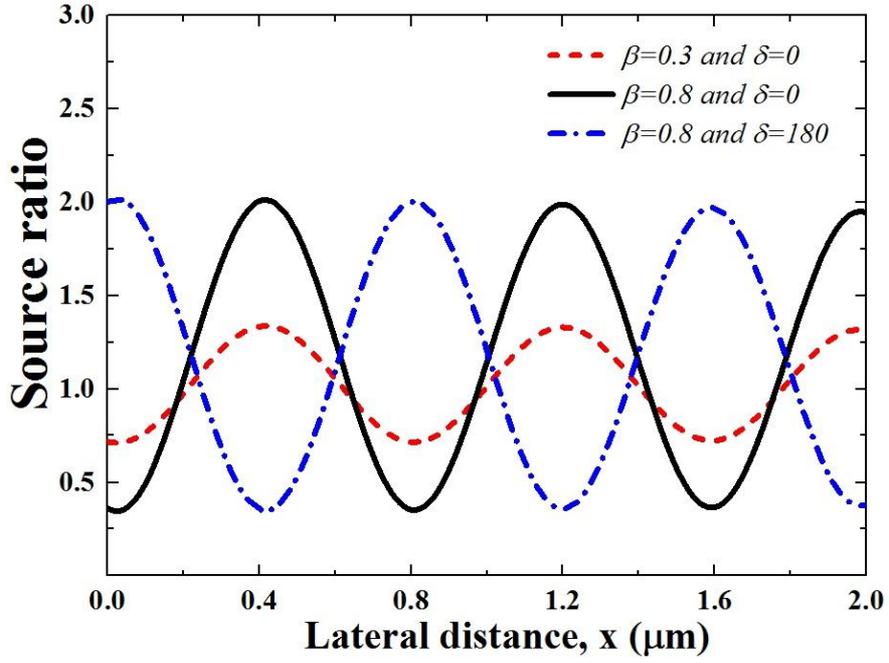

**Figure B1.** Dependence of the source term ratio on the lateral distance from the step edge at different coupling efficiencies $\beta$ and the phases $\delta$ at the surface $(z=0, y=0)$ and the moment of the pulse maximum $t = t_0$ for the laser wavelength of 800 nm.

Two values of the parameter $\beta$ are considered in the Fig. B1. If we reduce $\beta$, the amplitude of the energy modulation decreases, according to the naming of $\beta$ as the coupling efficiency. In contrast, when we change the phase $\delta$, the amplitude of the source term does not change, but it shifts depending on the value of $\delta$. Therefore, we can conclude that the phase $\delta$ describes the shift of the source term profile along the lateral distance.

**References**

1. B.N. Chichkov, C. Momma, S. Nolte, F. von Alvensleben, A. Tünnermann, Femtosecond, picosecond and nanosecond laser ablation of solids, Appl. Phys. A 63 (1996) 109-115.

2. R.R. Gattass, E. Mazur, Femtosecond laser micromachining in transparent materials, Nat. Photonics 2 (2008) 219-225.

3. A.Y. Vorobyev, C. Guo, Direct femtosecond laser surface nano/microstructuring and its applications, Laser Photonics Rev. 7 (2013) 385–407.

4. M.V. Shugaev, C. Wu, O. Armbruster, A. Naghilou, N. Brouwer, D.S. Ivanov, T.J.-Y. Derrien, N.M. Bulgakova, W. Kautek, B. Rethfeld, and L.V. Zhigilei, Fundamentals of ultrafast laser – material interaction, MRS Bull. 41 (2016) 960-968.

5. B. Rethfeld, D.S. Ivanov, M.E. Garcia and S.I. Anisimov, Modelling ultrafast laser ablation, J. Phys. D: Appl. Phys. 50 (2017) 193001.

6. P. Kahl, D. Podbiel, C. Schneider, A. Makris, S. Sindermann, C. Witt, D. Kilbane, M. Horn-von Hoegen, M. Aeschlimann, F. Meyer zu Heringdorf, Direct Observation of Surface Plasmon Polariton Propagation and Interference by Time-Resolved Imaging in Normal-Incidence Two Photon Photoemission Microscopy, Plasmonics 13 (2018) 239–246.

7. R.L. Olmon, B. Slovick, T.W. Johnson, D. Shelton, S.-H. Oh, G.D. Boreman, and M.B. Raschke, Optical dielectric function of gold, Phys. Rev. B 86 (2012) 235147.

8. E. Ozbay, Plasmonics: Merging Photonics and Electronics at Nanoscale Dimensions, Science 311 (2006) 189-193.

9. H.A. Atwater, The promise of plasmonics, Scientific American 296 (2007) 56-63.

10. M. Kauranen, A.V. Zayats, Nonlinear plasmonics, Nat. Phys. 6 (2012) 737–748.

11. N. Liu, F. Wen, Y. Zhao, Y. Wang, P. Nordlander, N.J. Halas, A. Aluì, Individual Nanoantennas Loaded with Three-Dimensional Optical Nanocircuits, Nano Lett. 13 (2013) 142-147.

12. D.S. Ivanov, V.P. Lipp, A. Blumenstein, F. Kleinwort, V.P. Veiko, E. Yakovlev, V. Roddatis, M.E. Garcia, B. Rethfeld, J. Ihlemann, and P. Simon, Experimental and Theoretical Investigation of
18